\documentclass[conference]{IEEEtran}
\IEEEoverridecommandlockouts

\usepackage{multirow}
\usepackage{float}
\usepackage{subfigure}
\usepackage{enumerate}
\usepackage{enumitem}
\usepackage{stfloats}
\usepackage{hyperref}
\usepackage{breakurl}
\usepackage{diagbox}
\usepackage{balance}
\usepackage{titlesec}
\usepackage{verbatimbox}
\usepackage{caption}
\usepackage{graphicx}
\usepackage[linesnumbered,ruled,vlined]{algorithm2e}
\usepackage{amsmath,amssymb,amsfonts}
\usepackage{booktabs}
\usepackage{color}

\renewcommand{\baselinestretch}{1}

\def\line{\rule[2pt]{0.4cm}{0.07em}}
\setitemize[1]{itemsep=0pt,partopsep=1pt,parsep=\parskip,topsep=1pt}

\title{\huge Attentive\;Knowledge-aware\;Graph\;Convolutional\;Networks with\;Collaborative\;Guidance\;for\;Personalized\;Recommendation}

\begin{document}

\author{\IEEEauthorblockN{Yankai Chen\textsuperscript{1*},
Yaming Yang\textsuperscript{2}, 
Yujing Wang\textsuperscript{2},
Jing Bai\textsuperscript{2},
Xiangchen Song\textsuperscript{3}, and
Irwin King\textsuperscript{1}~\IEEEmembership{Fellow,~IEEE}}
\IEEEauthorblockA{\textsuperscript{1}Department of Computer Science and Engineering,
The Chinese University of Hong Kong, Hong Kong}
\IEEEauthorblockA{\textsuperscript{2}Microsoft Research Asia, China, \textsuperscript{3}Department of Machine Learning, Carnegie Mellon University, USA}
\IEEEauthorblockA{\{ykchen, king\}@cse.cuhk.edu.hk, \{yayaming, yujwang, jbai\}@microsoft.com, xiangchensong@cmu.edu}
}

\maketitle
\begingroup\renewcommand\thefootnote{*}
\footnotetext{This work is partially done during internship at Microsoft Research Asia.}
\endgroup


\begin{abstract}
To alleviate data sparsity and cold-start problems of traditional recommender systems (RSs), incorporating knowledge graphs (KGs) to supplement auxiliary information has attracted considerable attention recently. 
However, simply integrating KGs in current KG-based RS models is not necessarily a guarantee to improve the recommendation performance, which may even weaken the holistic model capability.
This is because the construction of these KGs is independent of the collection of historical user-item interactions; hence, information in these KGs may not always be helpful for recommendation to all users.

In this paper, we propose \textit{attentive \underline{K}nowledge-aware \underline{G}raph convolutional networks with \underline{C}ollaborative \underline{G}uidance for personalized \underline{R}ecommendation} (CG-KGR). 
CG-KGR is a novel knowledge-aware recommendation model that enables ample and coherent learning of KGs and user-item interactions, via our proposed \textit{Collaborative} \textit{Guidance} \textit{Mechanism}.
Specifically, CG-KGR first encapsulates historical interactions to interactive information summarization. 
Then CG-KGR utilizes it as guidance to extract information out of KGs, which eventually provides more precise personalized recommendation.
We conduct extensive experiments on four real-world datasets over two recommendation tasks, i.e., Top-K recommendation and Click-Through rate (CTR) prediction. The experimental results show that the CG-KGR model significantly outperforms recent state-of-the-art models by 1.4-27.0\% in terms of Recall metric on Top-K recommendation.





\end{abstract}




\section{Introduction}
Recommender systems (RSs) nowadays play an increasingly important role throughout E-commerce platforms, social networks, and commercial websites. 
A traditional recommendation method - \textit{collaborative filtering} (CF) - models user preferences of interests based on the similarity of users or items from the historical interactions. Recent proposed graph neural network based methods simulating the CF process~\cite{wang2019neural,sun2020multi} demonstrate the remarkable improvement over traditional matrix factorization (MF) models~\cite{koren2009matrix,hofmann2004latent,koren2008factorization,ChenYNP0020}. However, CF-based RS models usually suffer from the data sparsity issue and cold-start problems~\cite{ma2007effective,volkovs2017dropoutnet,KhawarZ19}.

To alleviate these issues, incorporating \textit{knowledge graphs} (KGs) as side information for RS models has recently attracted considerable attention~\cite{RippleNet,KGCN,KGAT,KGNNLS,CKAN}. Essentially, a KG is a heterogeneous graph, where nodes denote \textit{entities} (i.e., products or items with associated attributes and properties) and edges represent mutual \textit{relations} among these entities. 
Based on KGs and user-item interactions, an intuitive and popular solution is to: first model user-item interactions and KGs in the graph data structure, and then design graph-based models, e.g., \textit{graph convolutional networks} (GCNs), to directly capture their semantic relations and topological structures. 
Instead of relying on interactive data solely, with the rich relational information in KGs to compensate for the sparsity, KG-integrated RS methods have the potential to provide more precise and interpretable recommendation.

\begin{figure}[tp]
\includegraphics[width=0.49\textwidth]{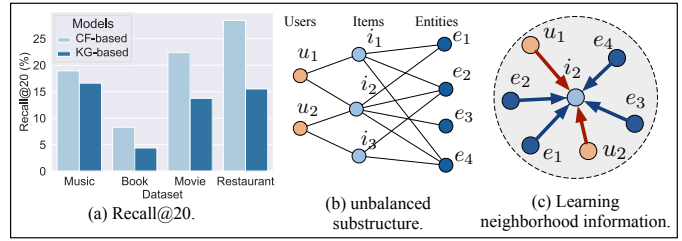}
\caption{Illustration of performance comparison, unbalanced graph substructure, and neighborhood information learning.}
\label{fig:comparison}
\end{figure}

\begin{figure}[tp]
\includegraphics[width=0.49\textwidth]{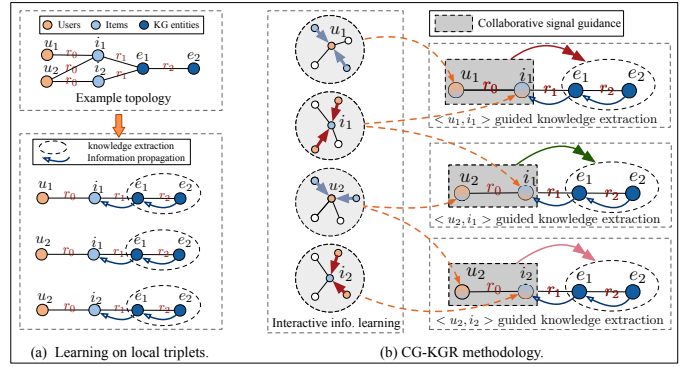}  
\caption{Illustration\,of\,knowledge\,extraction\,(best\,view\,in\,color).}
\label{fig:modelIllustration}
\end{figure}

\textbf{Major Motivation.}
Despite these promising benefits, simply integrating KGs in the RS models is not necessarily a guarantee to improve the recommendation performance.
As shown in Figure~\ref{fig:comparison}(a), some KG-based models even underperform the best traditional CF-based models on Recall metrics of Top-20 recommendation task (details of datasets and experimental analysis are referred to Section~\ref{sec:RQ1}).
The main reason is that these KGs are usually constructed independently of interaction data collection, where these two data sources may present unbalanced substructures, such as the different edge densities as shown in Figure\ref{fig:comparison}(b).
Furthermore, information in these KGs may not be all informative or helpful for recommendation to all users, which may even suppress the holistic model performance.
Thus, given the interaction data and KGs, the crux of boosting performance lies in better information extraction, by making sufficient and coherent use of them.
However, existing KG-aware RS models may fall short of satisfaction towards this goal mainly in three aspects:
\begin{itemize}[leftmargin=*]
\item \textbf{Explicit learning of interactive information is insufficient.}
Early studies~\cite{RippleNet,KGCN,KGNNLS} focus on encoding KGs triplets for knowledge supplement by propagating information among KG entities.  However, they do not explicitly propagate interactive information in the embedding learning, but only use the objective function to implicitly regularize the user-item embeddings.
Specifically, interaction data directly reveal the \textbf{users' item preferences} and \textbf{items' attracting groups}, which is important to enrich their latent profiles, as a basic assumption in recommendation is: users who accept certain items are likely to accept similar ones in the future (likewise for items).
Hence, without explicitly embedding historical interactive information, these methods may draw insufficient model learning for recommendation.

\item \textbf{Conventional methods simply mix the learning of interactive information and external knowledge.}
To enrich items' embeddings, recent work~\cite{KGAT,CKAN} purely relies on the graph topology to learn items' neighboring information, which may lead to unbalanced information aggregation from possibly unbalanced structures.
Furthermore, if external knowledge is redundant and not necessarily all helpful for recommendation, it may cause excessive collection of uninformative knowledge.
For instance, to learn item $i_2$'s embedding from Figure~\ref{fig:comparison} (b), conventional methods include interactive information and external knowledge at the same learning stage;
as shown in Figure~\ref{fig:comparison}(c), due to the high density in its KG side and low density in the interaction data side, $i_2$ obtains an unbalanced information summarization from external knowledge against its interactive counterpart.

\item \textbf{The effect of knowledge extraction from KGs for personalized recommendation is limited.}
Most work~\cite{RippleNet,KGAT,KGNNLS,CKAN} extracts knowledge by purely focusing on local KG structures.
As shown in Figure~\ref{fig:modelIllustration}(a), given the example topology in the upper half, conventional methods learn knowledge triplet $(e_1,r_2,e_2)$ to invariably diffuse the semantics within the interactive graph.
However, for different interaction pairs $<$$u_1$, $i_1$$>$, $<$$u_2$, $i_1$$>$, and $<$$u_2$, $i_2$$>$, such local knowledge can not adaptively match with different interests and backgrounds that $u_1$, $u_2$, $i_1$, and $i_2$ respectively have. 
As we just mentioned, target users and items tend to have diverse preferences and attracting groups.
Thus, in personalized recommendation, the informativeness of the same knowledge triplet actually varies and knowledge compositions for embedding enrichment should be dynamically adjusted.
In other words, the knowledge extraction phase should be \textbf{customized} accordingly.
Note that KGCN~\cite{KGCN} directly multiplies user embeddings with edge embeddings to weigh knowledge contribution. 
But simply using user and edge information is too general so that the distinct semantics of each triplet may be smoothed.
In a nutshell, current works may lack certain mechanisms to dynamically adjust the knowledge compositions for target users and items, leading to unsatisfactory personalized recommendation.

\end{itemize}

To address these limitations, we propose \textit{attentive} \textit{\underline{K}nowledge-aware} \textit{\underline{G}raph} \textit{convolutional} \textit{networks} \textit{with} \textit{\underline{C}ollaborative} \textit{\underline{G}uidance} \textit{for} \textit{personalized} \textit{\underline{R}ecommendation} (CG-KGR). 
CG-KGR approaches the problem for better personalized recommendation via GCN-based representation learning.
We distinguish the \textbf{interactive} \textbf{information} \textbf{summarization} from external \textbf{knowledge} \textbf{extraction}, via setting different learning strategies to provide balanced information aggregation from interaction data and KGs. 
Technically, we perform the sufficient learning of interactive data and coherent extraction of KG information as follows: 
\begin{itemize}[leftmargin=*]
\item In interactive information summarization, we profile users' preferences by visiting their interacted items and summarize the attracting groups of items by exploring their associated users.
By propagating the information back and forth between users and items, it simulates the \textit{collaborative filtering effect} for recommendation~\cite{ngcf}.
We explicitly embed such summarized information to make adequate learning of interaction data, which is crucial as the preparation for utilization in the following customized knowledge extraction.

\item To tailor knowledge extraction for better personalized recommendation, we propose a novel two-step mechanism called \textbf{Collaborative} \textbf{Guidance}.
As shown in Figure~\ref{fig:modelIllustration}(b): 
(1) given the target user-item pairs, .e.g., $<$$u_1$, $i_1$$>$, we first summarize their interactive information and encode it to the \textbf{collaborative} \textbf{guidance} \textbf{signal};
(2) then we immerse this signal into the knowledge extraction process that provides the early-matching functionality. 
Generally, the guidance signal contains the target user's information that serves as the \textbf{preference filtering} to mask the irrelevant information in KGs, and the target item's information that works as the \textbf{attraction grouping} to further highlight the important factors.
For instance, as shown in Figure~\ref{fig:modelIllustration}(b), different interaction pairs. e.g., $<$$u_1$, $i_1$$>$, have different guidance effects to knowledge triplet $(e_1,r_2,e_2)$, marked by the red, green, and pink arrows, which thus provide customized semantic enrichments to their corresponding embeddings.
\end{itemize}

Unlike traditional models~\cite{RippleNet,KGNNLS,KGAT,CKAN} adopting the straightforward KG semantics extraction for local triplets, e.g., $(h,r,t)$, CG-KGR develops the customized knowledge extraction based on the quintuplets $($$<$$u$,$i$$>$, $h$, $r$, $t$$)$. 
It provide a more \textbf{fine-grained} learning paradigm to enrich the embeddings of target users and items.
To summarize, our main contributions are as follows:

\begin{enumerate}[leftmargin=*]
\item We propose an end-to-end knowledge-aware recommendation model named CG-KGR.
CG-KGR applies the two-step information summarization called Collaborative Guidance Mechanism to distinguish the learning of interactive information from external knowledge, providing balanced semantic enrichment to target embeddings.
It develops a tailored knowledge extraction by binding user-item interactive information, which is a fine-grained learning paradigm to produce more precise personalized recommendation.

\item 
We conduct comprehensive experiments on four benchmarks over Top-K recommendation and Click-Through rate (CTR) prediction. 
The experimental results demonstrate that CG-KGR achieves improvements over baselines by 1.35-27.03\% of Recall@20 metric on Top-20 recommendation and 0.49-2.04\% of AUC metric on CTR prediction.

\end{enumerate}

\textbf{Organization.}
We first define the problem in Section~\ref{sec:pre} and then present the detailed methodology of the CG-KGR model in Section~\ref{sec:method}. In Section~\ref{sec:exp}, we report the experimental results on tasks of Top-K recommendation and CTR prediction. 
Finally, we review the related works in Section~\ref{sec:label} and conclude the paper in Section~\ref{sec:con}.

\section{Problem Formulation}
\label{sec:pre}

User-item interactions can be represented by a bipartite graph, i.e., $\{(u, r^*, i)|u$ $\in$ $\mathcal{U}$, $i$ $\in$ $\mathcal{I}\}$. $\mathcal{U}$ and $\mathcal{I}$ denote the sets of users and items, and $r^*$ generalizes all user-item interactions, e,g., \textit{browse}, \textit{click}, or \textit{purchase}, as one relation type between user $u$ and item $i$.
Moreover, we use $y_{u,i}$ $=$ $1$ to indicate there is an observed interaction between $u$ and $i$, otherwise $y_{u,i}$ $=$ $0$.
A KG is formally defined as $\{(e_1,r,e_2)|e_1,e_2$ $\in$ $\mathcal{E}$, $r$ $\in$ $\mathcal{R}\}$, denoting that relation $r$ connects entity $e_1$ and $e_2$. $\mathcal{E}$ and $\mathcal{R}$ represent the sets of entities and relations. The KG is used to provide side information such as item attributes and external knowledge for items, e.g., (\textit{La La Land}, \textit{ActedBy}, \textit{Ryan Gosling}).
Moreover, each item can be matched with an entity in the KG to achieve the alignment from items to some KGs entities, i.e., $\mathcal{I} \subseteq \mathcal{E}$,~\cite{KGAT,CKAN}. 
For ease of interpretation, unifying interaction data and item knowledge is thus defined as $\mathcal{G} = (\mathcal{E}', \mathcal{R}')$, where $\mathcal{E}'$ $=$ $\mathcal{E}$ $\cup$ $\mathcal{U}$ and $\mathcal{R}'$ $=$ $\mathcal{R}$ $\cup$$\{r^*\}$.  

\textbf{Notations.} We use bold lowercase, bold uppercase and calligraphy characters to denote vectors, matrices and sets. Non-bold ones are used to denote graph nodes or scalars. 

\textbf{Task description.} Given $\mathcal{G} = (\mathcal{E}', \mathcal{R}')$, the recommendation task studied in this paper is to train a RS model predicting the probability $\hat{y}_{u,i}$ that target user $u$ may adopt target item $i$.

\section{CG-KGR Model Methodology}
\label{sec:method}
We now present the details of our proposed CG-KGR model. Figure~\ref{fig:framework} depicts the model framework. 
In the following sections, we will demonstrate: (1) user-item interactive information summarization for guidance signal encoding; (2) knowledge extraction with collaborative guidance; (3) model prediction and optimization of CG-KGR, accordingly.

\subsection{\textbf{Interactive Information Summarization for Collaborative Signal Encoding}}
Interactive information summarization profiles user preferences and item attracting groups. 
Based on the summarized information, CG-KGR further encodes it to the guidance signal.
To explain the attentive information summarization, we start with the description of \textit{collaboration attention}.

\noindent\subsubsection{\textbf{Collaboration Attention.}}
\label{sec:CollaborativeAttention}
Given the \textbf{target interaction pair} ($u$, $i$) between user $u$ and item $i$, we compute the attentive weight $\pi(u, i)$:
\begin{equation}
\pi(u, i) = \boldsymbol{v}_u^T \boldsymbol{M}_{r^*} \boldsymbol{v}_i,
\end{equation}
where $\boldsymbol{v}_u, \boldsymbol{v}_i \in \mathbb{R}^d$ are the $d$-dimensional embeddings of $u$ and $i$. 
$\boldsymbol{M}_{r^*}\in \mathbb{R}^{d\times d}$ is the transformation matrix for relation $r^*$.
Then the normalized coefficients across all interactions from user $u$ can be computed by using the softmax function:
\begin{equation}
\label{eq:uiAtt}
\hat{\pi}(u,i) = \frac{\exp(\pi(u,i))}{\sum_{i' \in \mathcal{S}(u)} \exp(\pi(u,i'))}.
\end{equation}

Attention mechanism has been widely studied in many tasks~\cite{li2019improving,vaswani2017attention,gao2021open}. And our attention defined above depends on the embeddings $\boldsymbol{v}_u$, $\boldsymbol{v}_i$ and weight matrix $\boldsymbol{M}_{r^*}$. Generally, $\hat{\pi}(u,i)$ characterizes the informativeness of historical item neighbors, i.e., {\small $\mathcal{S}(u)$}, which enables user $u$ to adaptively incorporate information from his/her historical interacted items. 

\subsubsection{\textbf{User-centric Interactive Information Propagation.}}
\begin{figure*}
\includegraphics[width=1\textwidth]{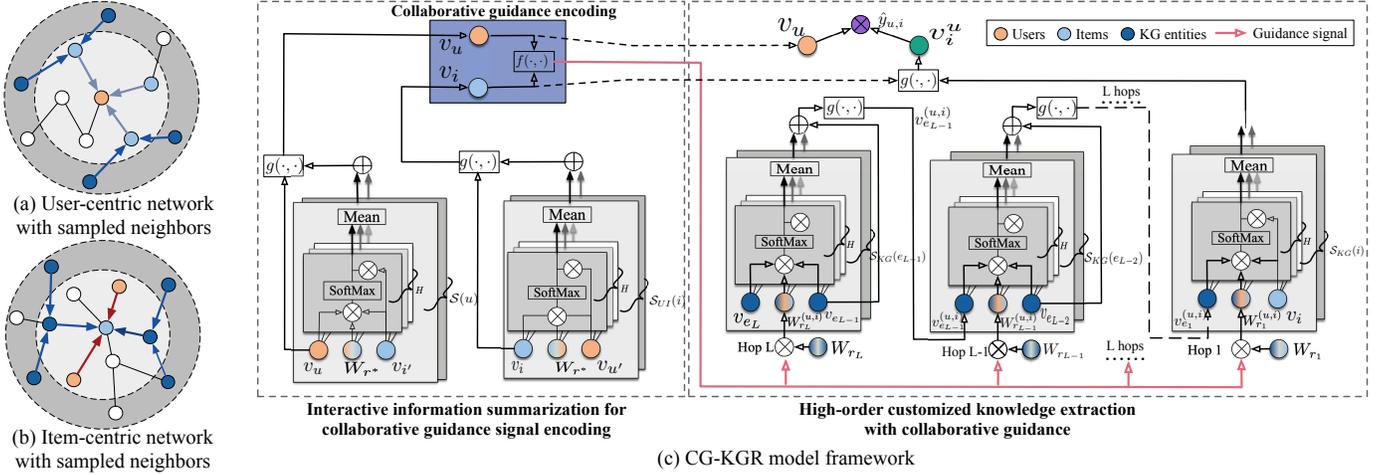}
\caption{(a) User-centric network (white nodes represent unsampled neighbors). 
(b) In the item-centric network, items are connected to users (brown) and KG entities (navy).
(c) Illustration of the proposed CG-KGR framework (best view in color).}
\label{fig:framework}
  \end{figure*}

As shown in Figure~\ref{fig:framework}(a), users are directly interacted with items. 
To profile the embedding of user $u$ by characterizing $u$'s historical item interactions, we compute the latent representation of $u$-centric network {\small $\mathcal{S}(u)$} as:
\begin{equation}
\label{eq:user_ngh}
\boldsymbol{v}^{}_{\mathcal{S}(u)} = \sum_{i' \in \mathcal{S}(u)}\hat{\pi}(u,i') \boldsymbol{v}_{i'}.
\end{equation}

Essentially, $\boldsymbol{v}_{\mathcal{S}(u)}$ is the linear combination of $u$'s neighbors in {\small $\mathcal{S}(u)$}. 
We extend our attention to \textit{averaging multi-head attention}~\cite{GAT} by taking the average of vanilla single attention mechanism that computed for $H$ times in parallel. 
Compared to single-head attention, it can further provide numerical stability for the learning process of self-attention in information propagation~\cite{GAT}.
Let $\hat{\pi}^{(h)}(u,i')$ denote the $h$-th normalized coefficient, and we redefine Equation~(\ref{eq:user_ngh}) as follows: 
\begin{equation}
\boldsymbol{v}^{}_{\mathcal{S}(u)} = \frac{1}{H}\sum_{h=1}^H \sum_{i' \in \mathcal{S}(u)}\hat{\pi}^{(h)}(u,i')  \boldsymbol{v}_{i'}.
\end{equation}

\subsubsection{\textbf{Item-centric Interactive Information Propagation.}}
Similarly, we compute the embedding of item $i$'s interactive neighbors, i.e., $\boldsymbol{v}_{\mathcal{S}_{UI}(i)}$, with multi-head attention as: 
\begin{equation}
\boldsymbol{v}^{}_{\mathcal{S}_{UI}(i)} = \frac{1}{H}\sum_{h=1}^{H} \sum_{u' \in \mathcal{S}_{UI}(i)}\hat{\pi}^{(h)}(i,u') \boldsymbol{v}_{u'},
\end{equation}
where $\hat{\pi}^{(h)}(i,u')$ is the $h$-th normalized coefficient that is computed by the collaboration attention.
Obviously, $\boldsymbol{v}_{\mathcal{S}_{UI}(i)}$ is calculated similarly to Equation~(\ref{eq:user_ngh}), as they share the same transformation matrix $\boldsymbol{M}_{r^*}$that generalizes the user-item relationship in the latent embedding space.

\textbf{Neighbor sampling.} Generally, feeding the whole graph to the graph convolutional networks suffers from highly computational overhead. Hence, neighbor sampling is often used to sample a sub-graph for efficient training~\cite{graphSage,RippleNet,KGCN,KGNNLS}.
This is particularly useful for web-scale recommender systems~\cite{ying2018graph}. 
We implement the dynamic neighbor sampling as a fixed-size random sampling in each training epoch, i.e., $\mathcal{S}(u)$ and $\mathcal{S}_{UI}(i)$, instead of using its full neighbors.  

\subsubsection{\textbf{Information Aggregation.}}
The next step is to aggregate propagated information with current information to iteratively update the node embeddings, by using the aggregation function $g(\cdot, \cdot)$: $\mathbb{R}^d$ $\times$ $\mathbb{R}^d$ $\rightarrow$ $\mathbb{R}^{d}$. 
In each iteration, we update embeddings, e.g., user $u$, as follows:
\begin{equation}
\boldsymbol{v}_u = g\Big(\boldsymbol{v}_u, \boldsymbol{v}_{\mathcal{S}(u)}\Big).
\end{equation}
Similarly, we can iteratively aggregate interactive information for items to update the embedding $\boldsymbol{v}_i$.


For aggregator selections, we utilize three types of aggregators to implement $g(\boldsymbol{v}_1, \boldsymbol{v}_2)$ of two inputs $\boldsymbol{v_1}$ and $\boldsymbol{v_2}$.
\begin{itemize}[leftmargin=*]
\item \textit{Sum Aggregator}~\cite{GCN} takes the summation of two inputs and conduct a nonlinear transformation:
\begin{equation}
g_{sum} = \sigma\Big(\boldsymbol{W}\cdot(\boldsymbol{v}_1 +  \boldsymbol{v}_2) + \boldsymbol{b}\Big).
\end{equation}
\item \textit{Concat Aggregator}~\cite{graphSage} concatenates two embeddings, followed by a nonlinear transformation:
\begin{equation}
g_{concat} = \sigma\Big(\boldsymbol{W}\cdot\big[\boldsymbol{v}_1 \big|\big| \boldsymbol{v}_2\big] + \boldsymbol{b}\Big).
\end{equation}
\item \textit{Neighbor Aggregator}~\cite{GAT} directly updates the output representation with the input embedding $\boldsymbol{v}_2$: 
\begin{equation}
g_{neighbor} = \sigma\Big(\boldsymbol{W}\cdot\boldsymbol{v}_2 + \boldsymbol{b}\Big).
\end{equation}
\end{itemize}

Here $\boldsymbol{W}$ and $\boldsymbol{b}$ are the trainable weight and bias. $\sigma$ is the nonlinear activation function such as ReLU.

\subsubsection{\textbf{Collaborative Guidance Signal Encoding.}}
Based on the updated embeddings of target user $u$ ($\boldsymbol{v}_u$) and target item $i$ ($\boldsymbol{v}_i$), we can encode them to the \textbf{guidance signal}.
In this paper, we implement three simple optional types of guidance signal encoders $f(\boldsymbol{v}_{u}, \boldsymbol{v}_{i})$: $\mathbb{R}^d$ $\times$ $\mathbb{R}^d$ $\rightarrow$ $\mathbb{R}^{d}$:
\begin{itemize}[leftmargin=*]
\item \textit{Sum Encoder} takes the summation of the inputs $\boldsymbol{v}_u$ and $\boldsymbol{v}_i$:
\begin{equation}
f_{sum} = \boldsymbol{v}_u + \boldsymbol{v}_i, 
\end{equation}

\item \textit{Pairwise-max Encoder} takes the element-wise maximum values of inputs $\boldsymbol{v}_u$ and $\boldsymbol{v}_i$:
\begin{equation}
f_{pmax} = {\rm pmax}(\boldsymbol{v}_{u}, \boldsymbol{v}_i) 
\end{equation}

\item \textit{Linear Combination Encoder} takes the linear combination of inputs $\boldsymbol{v}_u$ and $\boldsymbol{v}_i$ as:
\begin{equation}
f_{comb} = \alpha\boldsymbol{v}_u + (1-\alpha) \boldsymbol{v}_i, {\rm \ where \ } \alpha \in (0, 1).
\end{equation}

\end{itemize} 

Once we get the guidance signal that contains the information of user preferences and item attracting groups, CG-KGR employs it for customizing knowledge extraction from KGs to further enrich the node embeddings. 

\subsection{\textbf{Knowledge Extraction with Collaborative Guidance}}
As shown in Figure~\ref{fig:framework}(b), items are also associated with KG entities. 
Therefore, for target item $i$, CG-KGR needs to further extract the external knowledge from KG side, via incorporating the collaborative guidance signal, i.e., $f(\boldsymbol{v}_{u}, \boldsymbol{v}_{i})$. 

\subsubsection{\textbf{Knowledge-aware\,Attention\,with\,Collaborative\,Guidance.}}
In order to discriminate the importance of KG associations, given the KG triplet $(i,r,e)$ where $e$ is the KG entity that is associated with item $i$ by relation $r$, we define the quintuplet notation $\big($$<$$u$,$i$$>$,$i,r,e\big)$. 
This represents that $(i,r,e)$ is guided by the target pair $(u,i)$.  
We first get the general relation-specific matrix $\boldsymbol{M}_r \in \mathbb{R}^{d\times d}$, and then compute the customized transformation matrix with the guidance signal as:
\begin{equation}
\boldsymbol{M}_r^{(u,i)} = f(\boldsymbol{v}_{u}, \boldsymbol{v}_{i}) \odot  \boldsymbol{M}_r,
\end{equation}
where $f(\cdot, \cdot) \in \mathbb{R}^{d}$ is the guidance signal that seizes the interactive information between user $u$ and item $i$. $\odot$ represents the element-wise product with \textit{broadcast mechanism}.
Notice that $\boldsymbol{M}_r$ describes the relation $r$ in the ($d$$\times$$d$)-dimensional space. By fusing the guidance signal from the target pair $(u,i)$, $\boldsymbol{M}_r^{(u,i)}$ can simultaneously capture the relational representation of relation $r$ as well as the interactive information of $u$ and $i$, biasing the coefficient computation for $(i, r, e)$ as:
\begin{equation}
\label{eq:kgAtt}
\omega\big(\text{$<$}u, i\text{$>$}, i, r, e\big) = \boldsymbol{v}_i^T \boldsymbol{M}_r^{(u,i)} \boldsymbol{v}_e,
\end{equation}
and it can be normalized by adopting the softmax function as:
\begin{equation}
\label{eq:normal_kgAtt}
\resizebox{1\linewidth}{!}{$
\displaystyle
\hat{\omega}\big(\text{$<$}u, i\text{$>$}, i, r, e\big) = \frac{\exp\Big(\omega\big(\text{$<$}u, i\text{$>$}, i, r, e\big)\Big)}{\sum_{e' \in \mathcal{S}_{KG}(i)} \exp\Big(\omega\big(\text{$<$}u, i\text{$>$}, i, r, e'\big)\Big)},
$}
\end{equation}
where ${\mathcal{S}_{KG}(i)}$ represents item $i$'s neighbor set of KG entities.
Coefficient $\hat{\omega}\big($$<$$u$, $i$$>$, $i, r, e\big)$ measures the relative informativeness of KG triplet $(i,r,e)$ guided by the target pair ($u, i$). 

\subsubsection{\textbf{Knowledge Extraction and Aggregation.}}
Based on the target pair $(u,i)$, we explicitly anchor the guidance signal in the computation of the latent representation of $i$'s neighboring KG entities ${\mathcal{S}_{KG}(i)}$ as follows:
\begin{equation}
\label{eq:KGpropgation}
\boldsymbol{v}^{\left(u,i\right)}_{\mathcal{S}_{KG}(i)} = \frac{1}{H} \sum^H_{h=1} \sum_{e \in \mathcal{S}_{KG}(i)}\hat{\omega}^h\big(\text{$<$}u, i\text{$>$}, i, r, e\big)  \boldsymbol{v}_e.
\end{equation}

Likewise, we adopt the fixed-size random sampling for {\small $\mathcal{S}_{KG}(i)$}. 
After incorporating the tailored knowledge, CG-KGR can supplement the additional backgrounds for items as well as their related interactive information.
In each iteration of the embedding learning, we reuse $g$ to compute the embedding specifically with the guidance signal as:
\begin{equation}
\boldsymbol{v}_i^u = g\Big(\boldsymbol{v}_i, \boldsymbol{v}^{\left(u,i\right)}_{\mathcal{S}_{KG}(i)} \Big).
\end{equation}

\textbf{High-order knowledge extraction.}
As shown in Figure~\ref{fig:framework}(c), to further extract the high-order KG information and propagate it to items for better recommendation, we can stack more extraction hops in our proposed CG-KGR model. 
Centred at the user-item target pairs, e.g., $(u,i)$, we can \textit{random-walk-based} explore paths outwards, e.g., $i$ $\stackrel{r_1}{\line}$ $e_1$ $\stackrel{r_2}{\line}$ $\cdots$ $\stackrel{r_{L}}{\line}$ $e_{L}$, where $e_l\,$$\in\,$$\mathcal{E}$ and $r_l\,$$\in\,$$\mathcal{R}$. ($e_{l-1}$,\,$r_l$,\,$e_l$) is the $l$-th KG triplet on this path, where $L$ is the path length. Here integer $l\,$$>$$\,0$, and if $l\,$$=$$\,0$, $e_{l-1}\,$$=$$\,i$. 
To gather distant information along these paths, CG-KGR extends the aforementioned knowledge-aware attention in high-order knowledge extraction.
Concretely, guided by the target pair $(u,i)$ in the $l$-depth exploration, we get the neighbor set of KG entity $e_{l-1}$, i.e., ${\mathcal{S}_{KG}(e_{l-1})}$, and then formulate its embedding as: 
\begin{equation}
\label{eq:highOrder}
\boldsymbol{v}^{(u,i)}_{\mathcal{S}_{KG}(e_{l-1})} = \frac{1}{H} \sum^H_{h=1} \sum_{e_l \in \mathcal{S}_{KG}(e_{l-1})} \hat{\omega}^h\big(\text{$<$}u,i\text{$>$}, e_{l-1},r_l, e_{l}\big) \boldsymbol{v}_{e_l},
\end{equation}
where coefficient $\hat{\omega}^h\big(\text{$<$}u,i\text{$>$}, e_{l-1},r_l, e_{l}\big)$ can be normalized using the softmax function similarly as Equation~(\ref{eq:normal_kgAtt}), after the computation of unnormalized coefficient:
\begin{equation}
\label{eq:high_order_Att}
\omega\big(\text{$<$}u, i\text{$>$}, e_{l-1},r_l, e_{l}\big) = \boldsymbol{v}_{e_{l-1}}^T \boldsymbol{M}_{r_l}^{(u,i)} \boldsymbol{v}_{e_{l}}.
\end{equation}
Then we compute the embedding of KG entity $e_{l-1}$ anchored with the guidance signal accordingly:
\begin{equation}
\boldsymbol{v}^{(u,i)}_{e_{l-1}} = g\Big(\boldsymbol{v}_{e_{l-1}}, \boldsymbol{v}^{(u,i)}_{\mathcal{S}_{KG}(e_{l-1})} \Big),
\end{equation}
where $\boldsymbol{v}_{e_{l-1}}$ is the unique embedding of entity $e_{l-1}$, memorizing $e_{l-1}$'s original information. 
Please notice that knowledge triplet ($e_{l-1}, r_l, e_l$) is originally explored from item $i$, which means if $l$ $=$ $1$, {\small $\boldsymbol{v}^{(u,i)}_{\mathcal{S}_{KG}(e_{l-1})}$ $=$ $\boldsymbol{v}^{(u,i)}_{\mathcal{S}_{KG}(i)}$}. 
This is the embedding of $i$'s first-order entity neighbors defined in Equation~(\ref{eq:KGpropgation}). 

High-order knowledge extraction also relies on the \textbf{neighbor} \textbf{sampling} to generate a \textit{graph node flow}, which in essence is a multi-hop sub-graph where edges live in the consecutive hops. 
In each hop of KG exploration, we conduct fixed-size random sampling to collect the KG entities.
Based on the high-order knowledge extraction with collaborative guidance, item embeddings can be further enriched, which thus boosts the final recommendation performance. 

\textbf{Pseudocodes of the CG-KGR model.} 
We attach the pseudocodes of CG-KGR in Algorithm~\ref{alg:CG-KGR}.
As illustrated in Algorithm~\ref{alg:CG-KGR}, we first conduct interactive information summarization for guidance signal encoding (lines 2-9).
For the customized knowledge extraction, guided by the collaborative signal, we iteratively propagate the $l$-hop KG information from $l=L$ to $l=1$ (lines 10-14). 
Then on the $1$-hop subgraph, the condensed KG information is further aggregated to finally enrich item $i$'s representation (line 14). 
Please notice that we use {\small$\mathcal{S}_{KG}(i)^{(l)}$} to represent $i$'s $l$-hop neighbors.
$0$-hop neighbor of item $i$ is $i$ itself (line 11), so that if $l = 1$, $e = i$ and $\boldsymbol{v}^{(u,i)}_e = \boldsymbol{v}^u_i$ (line 14).
$B$ denotes the batch size.

\begin{algorithm}[h]
\small
\caption{CG-KGR algorithm}
\label{alg:CG-KGR}
\LinesNumbered  
\KwIn{User-item interactions $\{\mathcal{U}, \mathcal{I}\}$ and KG $\{\mathcal{E}, \mathcal{R}\}$; trainable parameters $\Theta$: $\{\boldsymbol{v}_u\}_{u\in\mathcal{U}}$, $\{\boldsymbol{v}_i\}_{i\in\mathcal{I}}$, $\{\boldsymbol{v}_e\}_{e\in\mathcal{E}}$, $\{\boldsymbol{M}_r\}_{r\in\mathcal{R}'}$, $\{\boldsymbol{W}_j, \boldsymbol{b}_j\}_{j=0}$; 
hyper-parameters: $d$, $L$, $B$, $H$, $\eta$, $\lambda$, $f(\cdot)$, $g(\cdot)$, $\sigma(\cdot)$. }
\KwOut{Prediction function $\mathcal{F}(u,i|\Theta, \mathcal{U}, \mathcal{I}, \mathcal{E}, \mathcal{R})$} 
\While{\rm{CG-KGR not converge}}{
    \For{$(u,i) \in \{\mathcal{U}, \mathcal{I}\}$ \rm{that} $y_{u,i}=1$}{
        $\mathcal{S}(u) \gets$ \texttt{Sample\_neighbor($u,1, \{\mathcal{U}, \mathcal{I}\}$)}; \\
        $\mathcal{S}_{UI}(i) \gets$ \texttt{Sample\_neighbor($i, 1, \{\mathcal{U}, \mathcal{I}\}$)};\\
        $\mathcal{S}_{KG}(i) \gets$ \texttt{Sample\_neighbor($i, L, \{\mathcal{E}, \mathcal{R}\}$)}; \\
        $\boldsymbol{v}_{\mathcal{S}(u)}, \boldsymbol{v}_{\mathcal{S}_{UI}(i)}$$\gets$summarize interactive information; \\
        $\boldsymbol{v}_u \gets g(\boldsymbol{v}_u, \boldsymbol{v}_{\mathcal{S}(u)})$;
        $\boldsymbol{v}_i \gets g(\boldsymbol{v}_i, \boldsymbol{v}_{\mathcal{S}_{UI}(i)})$; \\
        $f(\boldsymbol{v}_u, \boldsymbol{v}_i) \gets$ guidance signal encoding;\\ 
        $f$
        \For{$l = L, \cdots, 1$}{
          \For{$e \in$ \rm{($l$-$1$)-hop neighbor of } $i$ \text{in} $\mathcal{S}_{KG}(i)$}{
              $\mathcal{S}_{KG}(e) \gets$ $e$'s neighbors in $\mathcal{S}_{KG}(i)^{(l)}$; \\
              $\boldsymbol{v}^{(u,i)}_{\mathcal{S}_{KG}(e)} \gets$ extract KG information guided by the collaborative signal $f(\boldsymbol{v}_u, \boldsymbol{v}_i)$; \\
              $\boldsymbol{v}^{(u,i)}_e \gets g(\boldsymbol{v}_e, \boldsymbol{v}^{(u,i)}_{\mathcal{S}_{KG}(e)})$; \\

          }
        }
      $\hat{y}_{u,i} \gets$ compute estimated matching score;\\
      $\mathcal{L} \gets $ compute loss and optimize CG-KGR model;\\ 
    }
}
\KwRet $\mathcal{F}$.\\
\vspace{0.05in}
\SetKwFunction{FMain}{Sample\_neighbor($x, L, \mathcal{G}'$)}
\SetKwProg{Fn}{Function}{:}{\KwRet}
\Fn{\FMain}{
    $\mathcal{S}^{(0)} \gets x$; \\
    \For{$l = 1, \cdots, L$}{
        \For{$y \in \mathcal{S}^{(l-1)}$}{
              $\mathcal{S}^{(l)} \gets \mathcal{S}^{(l)} \cup \{z |\text{sampled neighbors of } y \text{ in } \mathcal{G}'\}$;
        }
    }
  \KwRet $\{\mathcal{S}^{(i)}\}^L_{i=0}$.
}
\end{algorithm}

\textbf{Time complexity analysis.}  
Let $c$ and $Y$ denote the number of epochs and user-item interactions, respectively. 
$\alpha$ is the average time cost of basic vector operations.
The holistic training time cost is $O\big(\alpha$$\cdot$$c$$\cdot$$Y$$\cdot$({\small$|\mathcal{S}_{KG}(i)|^L$}$+${\small$|\mathcal{S}_{UI}(i)|$}$+${\small$|\mathcal{S}(u)|)\big)$}.
In this paper, as we will present later, for four benchmarks, we have $c$ $\leq$ $10$; the sampling size for all nodes is no more than $16$.
Although the theoretical time complexity is exponential to $L$, in our work, $L$ $\leq$ $2$. 
This is because stacking too many extraction hops may incur performance detriment, the main cause of which lies in the well-known \textit{over-smoothing}~\cite{li2019deepgcns,li2018deeper} problem, i.e., vanishing gradient problem that leads to features of graph nodes converging to the same values. 
As we will show in Section~\ref{sec:time}, compared to recent state-of-the-art KG-aware models stacking limited hops ($L\leq 2$), CG-KGR is comparably efficient in practice.

\subsection{\textbf{Model Prediction and Optimization}}
\textbf{Model prediction.} 
In many embedding-based models, \textit{inner product} is widely adopted mainly for its simple but effective modeling of user-item interactions at the online matching stage.
During the ranking stage, items with top scores $\hat{y}_{u,i}$ are selected for recommendation to $u$. 
In this work, based on the learned embeddings of target user-item pair ($u,i$), we use it to directly estimate their matching score as: 
\begin{equation}
\setlength{\abovedisplayskip}{5pt}
\setlength{\belowdisplayskip}{5pt}
\hat{y}_{u,i} = \boldsymbol{v}_u^T  \boldsymbol{v}_i^u.
\end{equation}

\textbf{Model optimization.}
Let $\mathcal{Y}^+_u$ denote the positive interacted item set of user $u$, i.e., $\hat{y}_{u,i} = 1$, and $\mathcal{Y}^-_u$ represent the corresponding negative sampling set, i.e., $\hat{y}_{u,i} = 0$. 
To effectively optimize CG-KGR for training, in this paper, we set $|\mathcal{Y}^+_u|$ $=$ $|\mathcal{Y}^-_u|$. 
In each iteration of model training, we update $\mathcal{Y}^+_u$ and $\mathcal{Y}^-_u$ on the fly. Finally, the loss function is defined:

\begin{equation}
\setlength{\abovedisplayskip}{0pt}
\setlength{\belowdisplayskip}{0pt}
\mathcal{L} = \sum_{u\in \mathcal{U}} \Big( 
\sum_{i\in \mathcal{Y}^+_u} \mathcal{J}(y_{u,i}, \hat{y}_{u,i}) - \sum_{i\in \mathcal{Y}^-_u} \mathcal{J}(y_{u,i}, \hat{y}_{u,i}) \Big) + \lambda ||\Theta||_2^2.
\end{equation}


where $\mathcal{J}$ denotes the cross-entropy loss term, $\Theta$ is the set of trainable model parameters and embeddings, and $||\Theta||_2^2$ is the $L$2-regularizer parameterized by $\lambda$ to avoid over-fitting.

\section{Experiments}
\label{sec:exp}
We evaluate the CG-KGR model on the tasks of Top-K recommendation and Click-Through rate (CTR) prediction, to answer the following research questions:
\begin{itemize}[leftmargin=*]
\item \textbf{RQ1.} How does our proposed CG-KGR perform compared to the state-of-the-art recommendation methods?

\item \textbf{RQ2.} How is the time-efficiency of CG-KGR and other baselines in model training? 

\item \textbf{RQ3.} How does our proposed Collaborative Guidance Mechanism affect CG-KGR model performance? 

\item \textbf{RQ4.} What is the effect of each model component?

\item \textbf{RQ5.} How do different hyper-parameter settings affect CG-KGR model performance?
\end{itemize}

\subsection{\textbf{Dataset}}
To evaluate the effectiveness of CG-KGR, we directly utilize the following four open datasets (including the interactive data and corresponding KGs) for music, book, movie, and restaurant recommendations, respectively. 
Due to the diversity in the domain, data size, and distribution, all these four benchmarks are widely evaluated in recent works~\cite{RippleNet,KGCN,KGNNLS,CKAN}. The first three datasets are publicly accessible and the last one is contributed by Meituan-dianping Inc.~\cite{KGNNLS}. The statistics of the four datasets are summarized in Table~\ref{tab:datasets}.

\begin{itemize}[leftmargin=*]
\item \textbf{Last-FM (Music)}\footnote{\scriptsize\url{https://grouplens.org/datasets/hetrec- 2011/}} is a dataset of listening history collected by Last.fm music website. Musical tracks are viewed as items, and it consists of listening information from a set of nearly two thousand users. The corresponding KG contains 9,366 entities, 15,518 KG triplets, and 60 relation types.

\item \textbf{Book-Crossing (Book)}\footnote{\scriptsize\url{http://www2.informatik.uni-freiburg.de/~cziegler/BX/}} is a dataset of book ratings in Book-Crossing Community. Its related KG contains 77,903 entities, 151,500 triplets, and 25 relation types.

\item \textbf{MovieLens-20M (Movie)}\footnote{\scriptsize\url{https://grouplens.org/datasets/movielens/}} is a widely adopted benchmark for movie recommendation. It contains about 20 million ratings on MovieLens. 102,569 entities, 499,474 triplets, and 32 relation types are included in the corresponding KG.

\item \textbf{Dianping-food (Restaurant)} is a commercial dataset from Dianping.com\footnote{\scriptsize\url{https://www.dianping.com/}} consisting of over 10 million diverse interactions, e.g., clicking, saving, and purchasing, between about 2 million users and 1 thousand restaurants. Related KG owns 28,115 entities, 160,519 triplets, and 7 types of the relation.
\end{itemize}  

\begin{table}[h]
\centering
\caption{Statistics of datasets.}
\label{tab:datasets}
\setlength{\tabcolsep}{3mm}{
\begin{tabular}{c | c  c  c  c}
\toprule
            & Music & Book  & Movie     & Restaurant\\
\midrule
\midrule
  \# users     & {1,872} & {17,860}  & {138,159}   & {2,298,698}\\
  \# items       & {3,846} & {14,967}  & {16,954}    & {1,362}\\
  \# interactions & {42,346}  & {139,746} & {13,501,622}  & {23,416,418}\\
\midrule
  \# entities   & {9,366} & {77,903}  & {102,569}   & {28,115}\\
  \# relations    & {60}    & {25}    &   {32}    & {7}\\
  \# KG triplets   & {15,518}  & {151,500} & {499,474}   & {160,519} \\
 
\bottomrule
\end{tabular}}
\end{table}

\subsection{\textbf{Baselines}}
We include two streams of competing methods: traditional CF-based methods (BPRMF, NFM), and KG-based methods. 
In KG-based methods, there are two main related types: regularization-based methods (CKE and KGAT), and propagation-based methods (RippleNet, KGCN, KGNN-LS, CKAN). 

\begin{itemize}[leftmargin=*]
\item \textbf{BPRMF}~\cite{BPRMF} is a classical CF-based method that performs matrix factorization, optimized by the Bayesian personalized ranking optimization criterion.

\item \textbf{NFM}~\cite{NFM} is a neural factorization machine baseline for recommendation without KGs involved.

\item \textbf{CKE}~\cite{CKE} is a representative regularization-based recommendation method. CKE exploits semantic embeddings learned from TransR~\cite{TransR} with information from structural, textual, and visual domains to subsume matrix factorization under a unified Bayesian framework.

\item \textbf{KGAT}~\cite{KGAT} is a representative regularization-based model that collectively refines embeddings of users, items, and KG entities via jointly training interaction embeddings and KG embeddings. As suggested in~\cite{KGAT}, we use pre-trained embeddings from BPRMF to initialize KGAT. 

\item \textbf{RippleNet}~\cite{RippleNet} is a state-of-the-art propagation-based model. RippleNet uses a memory-like network to propagate user preferences towards items by following paths in KGs.

\item \textbf{KGCN}~\cite{KGCN} is another state-of-the-art propagation-based method that extends spatial GCN approaches to the KG domain. By aggregating high-order neighborhood information selectively and biasedly, both structure information and semantic information of the KG can be learned to capture users' potential interests.

\item \textbf{KGNN-LS}~\cite{KGNNLS} is a classical propagation-based method that applies graph neural network architecture to KGs with label smoothness regularization for recommendation.

\item \textbf{CKAN}~\cite{CKAN} is the latest state-of-the-art propagation-based method employing a heterogeneous propagation strategy to encode diverse information for better recommendation.
\end{itemize}

\subsection{\textbf{Experiment Setup}}
\label{sec:exp_setup}
To present reproducible and stable experimental results, we randomly split each dataset five times into training, evaluation, and test sets with the ratio of 6:2:2.
In our evaluation, we consider two recommendation tasks: (1)  Top-K recommendation and (2) Click-Through rate (CTR) prediction. 
\begin{itemize}[leftmargin=*]
\item In Top-K recommendation, we apply the trained model to rank $K$ items for each user with the highest predicted scores, i.e., $\hat{y}_{u,i}$. We choose two widely-used evaluation protocols Recall@$K$ and NDCG@$K$ to evaluate Top-K recommendation capability of CG-KGR model. 

\item For the CTR prediction, we first use sigmoid function to rescale $\hat{y}_{u,i}$, and then assign the click rate to 1 or 0 determined by the rescaled $\hat{y}_{u,i}$ with the threshold 0.5.
We adopt AUC and F1 as the evaluation metrics.
\end{itemize}


We implement the CG-KGR model under Python 3.7 and TensorFlow 1.14.0 with non-distributed training. 
The experiments are run on a Linux machine with a NVIDIA T4 GPU, 4 Intel Cascade Lake CPUs, 16 GB of RAM.
For all the baselines, we follow the official hyper-parameter settings from original papers or as default in corresponding codes. 
For methods lacking recommended settings, we apply a grid search for hyper-parameters.
The embedding size is searched in \{$8$, $16$, $32$, $64$, $128$\}. 
The learning rate $\eta$ is tuned within \{$10^{-3}, 5\times10^{-2}, 10^{-2}, 5\times10^{-1}$\}. We initialize and optimize all models with default Xavier initializer~\cite{Xavier} and Adam optimizer~\cite{Adam}.

\begin{table*}[]
\centering
  \caption{Average results of Top@20 recommendation task. Underlines indicate models with the second-best performance. Bolds denote the empirical improvements against second-best (underline) models, and $*$ denotes scenarios where a Wilcoxon signed-rank test indicates a statistically significant improvement over the second-best models under 95\% confidence level. }
  \label{tab:top20}
  \setlength{\tabcolsep}{1.6mm}{
  \begin{tabular}{c|c c|c c|c c|c c}
    \toprule
    \multirow{2}*{Model} & \multicolumn{2}{c|}{Music} & \multicolumn{2}{c|}{Book} & \multicolumn{2}{c|}{Movie} & \multicolumn{2}{c}{Restaurant} \\
               ~ & Recall@20(\%) & NDCG@20(\%) & Recall@20(\%) & NDCG@20(\%) & Recall@20(\%) & NDCG@20(\%) & Recall@20(\%) & NDCG@20(\%) \\

    \midrule
    \midrule
    
    BPRMF & {16.84} $\pm$ 3.86 & \ \ {8.75}  $\pm$ 1.94 & {4.67}  $\pm$ 0.87 & {2.80}  $\pm$ 0.43 & {20.48}  $\pm$ 1.57 & {15.77}  $\pm$ 0.91 & {19.90}  $\pm$ 3.02 & {10.79}  $\pm$ 2.03\\
    
    NFM & {11.51} $\pm$ 4.24 &\ \  {4.96} $\pm$ 2.18 & {3.93} $\pm$ 2.16 & {2.17} $\pm$ 1.49 & {19.79} $\pm$ 3.34 & {14.28} $\pm$ 1.14 & {23.85} $\pm$ 3.85 & {12.48} $\pm$ 2.98 \\
    
    CKE & {17.15} $\pm$ 5.30 & \ \ {8.82} $\pm$ 2.55 & {4.38}  $\pm$ 0.96 & {2.24} $\pm$ 4.20 & {21.52} $\pm$ 1.21 & {15.73} $\pm$ 1.31 & {22.24} $\pm$ 3.05 & {12.09} $\pm$ 1.51\\
    
    RippleNet & {16.61} $\pm$ 3.96 &\ \  {8.14} $\pm$ 1.57 & {7.12} $\pm$ 2.07 & {5.09} $\pm$ 1.68 & {13.74} $\pm$ 2.63 & {\ \,9.77} $\pm$ 1.70 & {21.20} $\pm$ 4.12 & {10.99} $\pm$ 1.99\\
    
    KGNN-LS & {17.73} $\pm$ 2.35 &\ \  {9.11} $\pm$ 1.08 & \underline{{8.51}} $\pm$ 2.21 & \underline{{6.06}} $\pm$ 1.66 & {20.20} $\pm$ 1.04 & {15.49} $\pm$ 1.30 & {15.52} $\pm$ 4.87 &  {\ \,7.92} $\pm$ 2.96\\
    
    KGCN & {18.25} $\pm$ 2.53 & \ \ {9.73} $\pm$ 1.54 & {7.85} $\pm$ 2.89 & {5.93} $\pm$ 2.25 & {19.24} $\pm$ 3.18 & {13.87} $\pm$ 1.55 & {19.03} $\pm$ 3.02 & \ \,{9.34} $\pm$ 1.51\\
    
    KGAT & {18.22} $\pm$ 4.30 & \ \ {9.31} $\pm$ 2.49 & {5.34}  $\pm$ 0.61 & {3.01} $\pm$ 0.79 & {\underline{21.80}} $\pm$ 0.77 & {\underline{16.81}} $\pm$ 1.05 & {15.57} $\pm$ 2.37 & {\ \,7.67} $\pm$ 1.72\\
    
    CKAN & {\underline{20.78}} $\pm$ 3.20 & {\,\underline{11.94}} $\pm$ 1.89 & {6.19} $\pm$ 1.14 & {3.47} $\pm$ 0.53  & {17.48} $\pm$ 1.73 & {12.48} $\pm$ 1.36 & {\underline{24.10}} $\pm$ 3.87 & {\textbf{13.33}} $\pm$ 2.04 \\
    
    \midrule
    \midrule 
    \textbf{CG-KGR} &{\textbf{21.06}} $\pm$ 3.52  & {\textbf{12.30}}$^*$ $\pm$ 1.82 & {\textbf{10.81}}$^*$ $\pm$ 3.59 & {\textbf{8.10}}$^*$ $\pm$ 2.71 & {\textbf{24.95}}$^*$ $\pm$ 1.91 & {\textbf{19.48}}$^*$ $\pm$ 1.02  & {\textbf{25.40}}$^*$ $\pm$ 4.88 & {\underline{12.55}} $\pm$ 1.64\\
    \% Gain & 1.35\%       & 3.02\%     & 27.03\%     & 33.66\%   & 14.45\%   & 15.88\%  & 5.39\%   & N/A   \\
    \bottomrule
  \end{tabular}}
\end{table*}

\begin{figure*}[]
\includegraphics[width=1\textwidth]{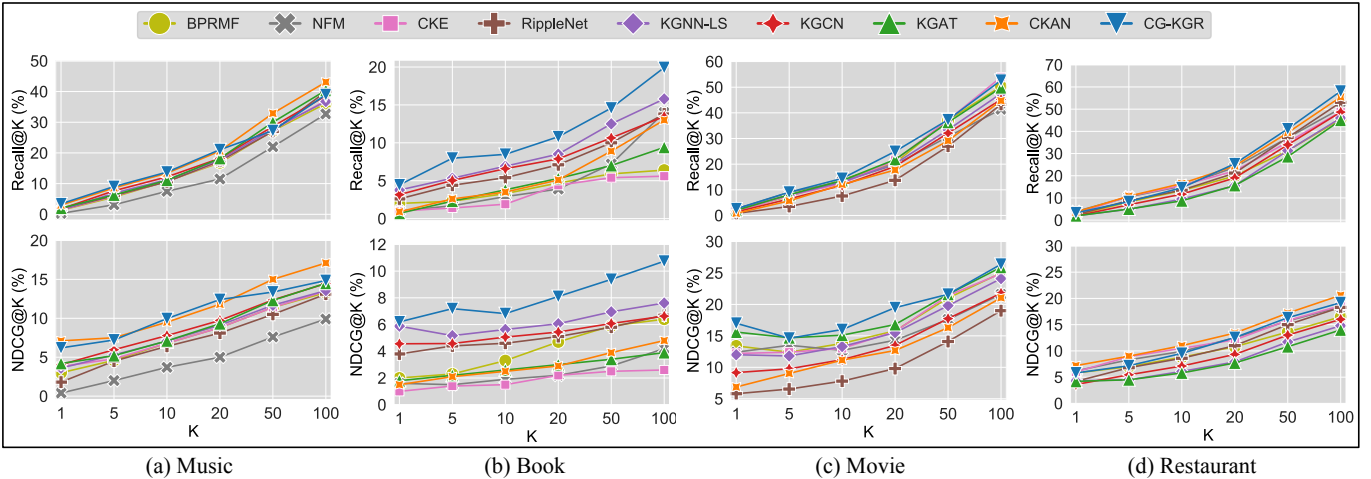}
\caption{Average results of Recall@$K$ and NDCG@$K$ in Top-K recommendation.}
\label{fig:topk} 
\end{figure*}

\subsection{\textbf{Performance Analysis (RQ1)}}
\label{sec:RQ1}
In this section, we present a comprehensive performance analysis between the CG-KGR model and all representative baselines in the tasks of Top-K recommendation and CTR prediction. 
For each task, we conduct Wilcoxon signed-rank tests~\cite {wilcoxon1992individual} to evaluate the significance of the improvement between the best performing model and the second-best model.

\subsubsection{\textbf{Top-K Recommendation.}}
We evaluate Top-K recommendation over four datasets by varying $K$ in \{1, 5, 10, 20, 50, 100\}. 
To achieve a more detailed comparison between CG-KGR and all baselines, we first summarize the results of Top@20 recommendation as well as corresponding standard deviations in Table~\ref{tab:top20}.
Then we curve the complete results of Recall@$K$ and NDCG@$K$ metrics of all baselines in Figure~\ref{fig:topk}. 
We have the following observation from the experimental results:
\begin{itemize}[leftmargin=*]

\item \textbf{Not all KG-based methods outperform traditional CF-based methods (without using KGs) on Top-K recommendation.}
As shown in Table~\ref{tab:top20} and Figure~\ref{fig:topk}, traditional CF-based methods BPRMF works slightly better than methods RippleNet on Music dataset and CKE on Book dataset, respectively. 
For the movie dataset, BPRMF outperforms most existing KG-based methods except CKE and KGAT.
For the largest dataset Restaurant, another CF-based method NFM, only underperforms CKAN but performs better than other KG-based methods. 
This phenomenon indicates that simply integrating KGs in recommendation is not necessarily a guarantee of performance improvement.
One possible reason to explain this is all these KG-based methods fully trust the information in KGs and do not conduct the tailored information extraction for a personalized recommendation. 
While in practice, information in KGs may not all be helpful.
So the key to boosting recommendation capability is to make sufficient and coherent use of KGs. In this paper, we propose CG-KGR as one solution to attain this goal.


\item\textbf{The results of Top-20 recommendation prove that the performance improvements of CG-KGR are statistically stable and significant.}
As shown in Table~\ref{tab:top20}, on Music, Book, Movie datasets, CG-KGR surpasses the state-of-the-art models \textit{w.r.t} Recall@20 and NDCG@20 by 1.35\%, 27.03\%, 14.45\%, and and 3.02\%, 33.66\%, 15.88\%, respectively. 
As for Restaurant dataset, CG-KGR achieves 5.39\% of improvement on Recall@20 but does not perform the best on NDCG@20. This is probably because, CG-KGR can make good prediction on retrieving the top 20 items from the candidate corpus but may not precisely estimate their specific relative item order for each user, since Restaurant contains the fewest items compared to other three datasets.
The standard deviations show that the results of our model are on the same level of stability as these state-of-the-art methods. 
Furthermore, Wilcoxon signed-rank tests verify that most improvements over the second-best model are statistically significant under 95\% confidence level. 

\item \textbf{As the value of K increases, CG-KGR consistently performs competitively compared to baselines.}
As shown in Figure~\ref{fig:topk}, compared to state-of-the-art models, CG-KGR consistently obtains superior performance when K varies on Book and Movie datasets and shows competitive performance on Restaurant dataset.
By explicitly propagating interactive information between users and items, CG-KGR is capable of learning latent representations of user preferences and item attracting groups from the historical interactions.
Furthermore, instead of directly integrating external knowledge without introducing internal interactive information, CG-KGR model applies the collaborative guidance mechanism. 
This mechanism collectively encodes historical interactions as guidance for the customized knowledge extraction, which is particularly useful in the personalized ranking task: Top-K recommendation. 

We observe that the performance curve of CG-KGR on Music dataset becomes flatter for larger K values than smaller ones.   
This is probably because the effect of knowledge supplement on Music dataset is relatively limited.
We use $\frac{\# KG \ triplets}{\# items}$ to measure the amortized volume of knowledge contribution to enrich item embeddings.
Obviously, a higher value usually indicates that the KG owns richer semantics to boost items' backgrounds.
While $\frac{\# KG \ triplets}{\# items}$ of Music dataset is 4.03, the other three datasets present the value of 10.12, 29.46, 117.86, respectively. 
Thus CG-KGR can perform much better on Book, Movie, and Restaurant datasets, by making sufficient use of rich semantics in these KGs to make precise item recalling from smaller K to larger one but presents limited performance, e.g., on Top-100 item recalling, over datasets with little knowledge contribution, e.g., Music dataset.


\end{itemize}

\subsubsection{\textbf{Click-Through Rate (CTR) Prediction.}}
\begin{table*}
\centering
  \caption{Average results of CTR prediction task. Underlines indicate models with the second-best performance. Bolds denote the empirical improvements against second-best (underline) models, and $*$ denotes scenarios where a Wilcoxon signed-rank test indicates a statistically significant improvement over the second-best models under 95\% confidence level.}
  \label{tab:ctr}
  \setlength{\tabcolsep}{1.6mm}{
  \begin{tabular}{c|c c|c c|c c|c c}
    \toprule
    \multirow{2}*{Model} & \multicolumn{2}{c|}{Music} & \multicolumn{2}{c|}{Book} & \multicolumn{2}{c|}{Movie} & \multicolumn{2}{c}{Restaurant} \\
               ~ & AUC(\%) & F1(\%) & AUC(\%) & F1(\%) & AUC(\%) & F1(\%) & AUC(\%) & F1(\%)\\

    \midrule
    \midrule
    BPRMF & {78.68} $\pm$ 0.32 & {71.64} $\pm$ 0.53 &  {60.53} $\pm$ 0.26 & {54.94} $\pm$ 0.56 & {97.54} $\pm$ 0.01 & {92.44} $\pm$ 0.02 & {84.32} $\pm$ 0.02 & {74.97} $\pm$ 0.02\\
    
    NFM & {78.03} $\pm$ 0.87 & {71.24} $\pm$ 1.05 & {72.07} $\pm$ 0.17 & {62.21} $\pm$ 2.99 & {96.64} $\pm$ 0.02 & {90.32} $\pm$ 0.47 & {87.25} $\pm$ 0.01 & {78.48} $\pm$ 0.70 \\
    
    CKE & {79.02} $\pm$ 0.50 & {71.25} $\pm$ 0.59 &  {61.07} $\pm$ 1.46 & {54.76} $\pm$ 0.01 & {97.65} $\pm$ 0.01 & {92.76} $\pm$ 0.02 & {83.93} $\pm$ 0.03 & {74.69} $\pm$ 0.04\\

    RippleNet & {80.37} $\pm$ 0.42 & {72.11} $\pm$ 0.71 & {71.81} $\pm$ 1.17 & {64.81} $\pm$ 0.52 & {97.63} $\pm$ 0.03 & {93.03} $\pm$ 0.06 & {87.58} $\pm$ 0.08 & {79.37} $\pm$ 0.07\\
    
    KGNN-LS & {77.82} $\pm$ 4.65 & {69.50} $\pm$ 3.33 & {68.96} $\pm$ 0.23  & {62.91} $\pm$ 0.37 & {97.88} $\pm$ 0.02 & {\underline{93.43}} $\pm$ 0.03 & {83.86} $\pm$ 0.57 & {77.40} $\pm$ 0.14\\
    
    KGCN & {79.07} $\pm$ 3.04 & {70.32} $\pm$ 2.20 &{67.09} $\pm$ 4.68 & {62.15} $\pm$ 2.22 & {97.69} $\pm$ 0.05 & {93.07} $\pm$ 0.09 & {84.97} $\pm$ 0.05 & {78.24} $\pm$ 0.07\\
    
    KGAT & {81.63} $\pm$ 0.69 & {\underline{74.29}} $\pm$ 0.76 & {68.21} $\pm$ 2.83 & {66.04} $\pm$ 4.09 & {\underline{97.94}} $\pm$ 0.01 & {93.25} $\pm$ 0.01 & {82.95} $\pm$ 0.53 & {76.02} $\pm$ 1.96\\
    
    CKAN & {\textbf{83.39}} $\pm$ 0.76 & {\textbf{75.94}} $\pm$ 0.85 & {\underline{74.38}} $\pm$ 0.43 & {\underline{66.73}} $\pm$ 0.48  & {97.17} $\pm$ 0.01 & {92.28} $\pm$ 0.01 & {\underline{87.84}} $\pm$ 0.02 & {\underline{80.19}} $\pm$ 0.03 \\
    
    \midrule
    \midrule
    \textbf{CG-KGR} &{\underline{83.00}} $\pm$ 0.68  & {73.74} $\pm$ 0.91 & {\textbf{75.78}}$^*$ $\pm$ 0.30 & {\textbf{67.14}}$^*$ $\pm$ 0.34 & {\textbf{98.42}}$^*$ $\pm$ 0.02 & {\textbf{94.38}}$^*$ $\pm$ 0.03 & {\textbf{89.63}}$^*$ $\pm$ 0.08  & {\textbf{81.88}}$^*$ $\pm$ 0.14\\
    \% Gain & N/A   &  N/A    &  1.88\%  & 0.61\%  & 0.49\%   & 1.02\%   & 2.04\%   & 2.11\%   \\
    \bottomrule
  \end{tabular}}
\end{table*}

Table~\ref{tab:ctr} summarizes the experimental results of CTR prediction task over all datasets.
Based on the results, we have the following observation and analysis.

\begin{itemize}[leftmargin=*]
\item \textbf{Our proposed model achieves effective and significant performance improvements \textit{w.r.t} AUC metric of CTR prediction.}
Concretely, CG-KGR improves the baselines on Book, Movie and Restaurant datasets \textit{w.r.t} AUC by 1.88\%, 0.49\% and 2.04\%, as well as F1 by 0.61\%, 1.02\%, and 2.11\%, with low variance respectively. 
Compared to the Top-K recommendation task, the performance gap between CG-KGR and baselines on CTR prediction is relatively smaller. 
This is because, while Top-K recommendation is a listwise ranking task that predicts the order of items to recommend; CTR prediction is essentially a pairwise classification task, which is actually easier. 
This means that the baselines can also perform well on CTR prediction and their performance gaps against CG-KGR are thus not that large. 
Moreover, we conduct the significant tests on CTR prediction. 
Based on the Wilcoxon signed-rank tests, our CG-KGR model evinces to improve CTR prediction performance significantly. 

\item \textbf{As for Music dataset, CG-KGR performs the second on AUC metric but does not show a prominent performance on F1 metric.}
As we have explained in the previous section, the value $\frac{\# KG \ triplets}{\# items}$ of Music dataset is the lowest.
This implies that, for our proposed CG-KGR, the effect of knowledge extraction to enrich the item embeddings is limited.
In addition, another possible explanation is that, after the normalization of predicted score $\hat{y}_{u,i}$, we simply set the threshold as 0.5 to determine whether item $i$ will be recommended to user $u$. 
With the limited and unbalanced distribution of positive and negative samples in Music, 0.5 may not be an appropriate threshold for binary classification. 
In contrast, AUC scores actually evaluate the model performance averaged over a whole range of thresholds, which, therefore, can better measure the classification capability of RS models under such circumstances.

\end{itemize}

\subsection{\textbf{Time Efficiency Comparison (RQ2)}}
\label{sec:time}
In this section, we study how time-efficient our CG-KGR and baselines are in model training. 
All methods run on the same aforementioned running environment without parallel training, and we use the default hyper-parameters that are reported in papers or official codes.
Table~\ref{tab:time} reports the average results of time cost per epoch, denoted by $\overline{t}$, and the numbers of epochs to reach the best performance, denoted by $\overline{be}$. Please notice that we apply the \textit{early stopping strategy} for all methods to prevent over-fitting, and the trigger condition is: the model performance is non-increasing for 10 consecutive epochs after the $\overline{be}$-th epoch.  
We can observe that:
\begin{itemize}[leftmargin=*]
\item CG-KGR requires close training time per epoch $\overline{t}$ with most of the efficient baselines (both CF-based and KG-based methods) on Music, Book, and Movie datasets.
In addition, CG-KGR runs faster than the latest KG-aware methods, i.e., KGAT, CKAN, on these datasets.
This shows the per-epoch training efficiency of CG-KGR model on the small to medium-sized datasets. 
On the largest Restaurant dataset, CG-KGR spends more training time per epoch to maximize the performance.
This is because of different model designs, other state-of-the-art methods obtain their best performance within no more than 2-hops of information propagation~\cite{RippleNet,CKAN}; while for our proposed CG-KGR model, it may include more neighbor samples to make sufficient learning for both collaborative guidance signal and further customized knowledge extraction. 
According to the aforementioned time complexity analysis, such time complexity is sustainable as well for large-size datasets.

\item {CG-KGR converges the fastest among all these baselines; thus the holistic training time is comparable with the latest state-of-the-art methods.} On Restaurant dataset, although CG-KGR needs more training time per epoch, it only requires about 2 epochs to converge. 
This means that the total training time of CG-KGR (i.e., 5,313.93$\times$2.2$\approx$1.1$\times$$10^4$(s)) is still competitive with some recent state-of-the-art works, e.g., CKAN costs 569.12$\times$11.4$\approx$6.5$\times$$10^3$(s), and KGAT costs 2,619.77$\times$2.4$\approx$6.3$\times$$10^3$(s).
Considering the performance improvements of CG-KGR on this dataset, we argue that the time cost is acceptable in practice.
\end{itemize}

\begin{table}[tbp]
\centering
\caption{Time cost (s) per epoch of model training.}
\label{tab:time}
\setlength{\tabcolsep}{1mm}{
\begin{tabular}{c |c c|c c|c c|c c}
\toprule
 \multirow{2}*{Model} & \multicolumn{2}{c|}{Music} & \multicolumn{2}{c|}{Book} & \multicolumn{2}{c|}{Movie} & \multicolumn{2}{c}{Restaurant} \\
               ~ & $\overline{t}$ & $\overline{be}$ & $\overline{t}$ & $\overline{be}$ & $\overline{t}$ & $\overline{be}$ & $\overline{t}$ & $\overline{be}$\\
\midrule
\midrule
  BPRMF     &{1.44}& {11.4}   & {5.93}& {47.4}   & {143.39}& {642.6}   & {163.97}& {2.6} \\
  NFM       &{9.38}& {87.8}   & {29.51}& {26.0}   & {112.27}& {5.8}   & {391.71}& {6.0} \\
  CKE       &{2.37}& {528.4}   & {13.18}& {108.8}  & {92.18}& {41.0}  & {109.15}& {476.6}  \\
  RippleNet &{4.28}& {8.6}   & {14.62}& {14.8}   & {1,393.94}& {8.0} & {2,564.01}& {4.2}\\
  KGNN-LS   &{1.43}& {5.6}   & {2.10}&  {6.0}  & {41.83}& {7.4}  & {92.91}&  {6.8} \\
  KGCN      &{1.23}& {5.4}   & {5.31}&  {8.4}  & {16.32}& {4.8}  & {58.10}& {3.6}  \\
  KGAT      &{13.40}& {207.2}  & {79.18}& {30.0}   & {371.04}& {2.2} & {2,619.77}& {2.4} \\
  CKAN      &{1.46}& {15.8}   & {3.18}& {10.4}   & {468.93}&  {18.2} & {569.12}& {11.4}  \\
\midrule
  CG-KGR    &{1.75}& {5.4} & {1.83}& {3.6}  & {321.83}& {4.2}  & {5,313.93}& {2.2} \\
\bottomrule
\end{tabular}}
\end{table}

\subsection{\textbf{Analysis of Collaborative Guidance Mechanism (RQ3)}}
In this section, we first conduct an ablation study to evaluate the effectiveness of collaborative guidance by masking information in the collaborative signal. 
Then we give a case study for visualization and evaluate the robustness of CG-KGR and baselines with corrupted information in Book dataset. 

\subsubsection{\textbf{Ablation Study of Collaborative Guidance Mechanism.}}

\begin{table}[ht]
\setlength{\abovecaptionskip}{0.4cm}
\setlength{\belowcaptionskip}{0cm}
\centering
\caption{Ablation study of Collaborative Guidance Mechanism on Top-20 recommendation (\%).}
\label{tab:collaborativeGuidance}

\setlength{\tabcolsep}{2mm}{
\begin{tabular}{c| c | c | c | c}
\toprule
  Dataset      & CG-KGR$_{NE}$ & CG-KGR$_{PF}$ & CG-KGR$_{AG}$   &  Best \\
\midrule
\midrule
       MS-R@20    & {19.22}{\ \,{\scriptsize(-8.74\%)}}  & {20.01}{\ \,{\scriptsize(-4.99\%)}}      & {19.29}{\ \,{\scriptsize(-8.40\%)}}          & {\textbf{21.06}} \\
       MS-N@20    & {10.78}{\,{\scriptsize(-12.36\%)}}  & {11.43}{\ \,{\scriptsize(-7.07\%)}}      & {10.92}{\,{\scriptsize(-11.22\%)}}        & {\textbf{12.30}} \\   
 \midrule
       BK-R@20    & {10.56}{\ \,{\scriptsize(-2.31\%)}}  & {10.29}{\ \,{\scriptsize(-4.81\%)}}      & {10.36}{\ \,{\scriptsize(-4.16\%)}}       & {\textbf{10.81}} \\
       BK-N@20    & {\ \,7.64}{\ \,{\scriptsize(-5.68\%)}}    & {\ \,7.87}{\ \,{\scriptsize (-2.84\%)}}      & {\ \,7.91}{\ \,{\scriptsize(-2.35\%)}}     & {\ \,\textbf{8.10}} \\   
 \midrule
       MV-R@20    & {23.76}{\ \,{\scriptsize(-4.77\%)}}  & {24.10}{\ \,{\scriptsize(-3.41\%)}}      & {24.28}{\ \,{\scriptsize(-2.69\%)}}         & {\textbf{24.95}} \\
       MV-N@20    & {18.67}{\ \,{\scriptsize(-4.16\%)}}  & {18.93}{\ \,{\scriptsize(-2.82\%)}}      & {19.24}{\ \,{\scriptsize(-1.23\%)}}       & {\textbf{19.48}} \\   
 \midrule
       RT-R@20    & {20.84}{\,{\scriptsize(-17.95\%)}}   & {23.36}{\ \,{\scriptsize(-8.03\%)}}      & {23.68}{\ \,{\scriptsize(-6.77\%)}}       & {\textbf{25.40}} \\
       RT-N@20    & {10.39}{\,{\scriptsize(-17.21\%)}}   & {11.94}{\ \,{\scriptsize(-4.86\%)}}      & {12.13}{\ \,{\scriptsize(-3.35\%)}}      & {\textbf{12.55}} \\   

\bottomrule
\end{tabular}}
\end{table}

We evaluate Collaborative Guidance Mechanism on the task of Top-K recommendation and report the results in terms of Recall@20 and NDCG@20 on four datasets that are respectively denoted by MS, BK, MV, and RT. 
Specifically, we set three variants of CG-KGR model.
CG-KGR$_{NE}$ is the model variant that simply encodes \underline{\textbf n}ode \underline{\textbf e}mbeddings (i.e., user and item embeddings and masking their historical interactive information) in the guidance signal.
We use CG-KGR$_{PF}$ to denote the model variant that conducts the \underline{\textbf p}reference \underline{\textbf f}iltering by only summarizing the users' historical interactions. 
CG-KGR$_{AG}$ denotes the variant that only explores the items' local structures for \underline{\textbf a}ttraction \underline{\textbf g}rouping.
Based on the results in Table~\ref{tab:collaborativeGuidance}, we have the following observations:
\begin{enumerate}[leftmargin=*]
\item Compared to our complete model implementation, variant CG-KGR$_{NE}$ presents a large performance decay ranging 2.31-17.95\% and 5.68-17.21\% \textit{w.r.t} Recall@20 and NDCG@20 over four datasets. 

\item Compared to CG-KGR$_{NE}$, partially using users' (CG-KGR$_{PF}$) or items' neighbor information (CG-KGR$_{AG}$) helps to boost the performance.
This demonstrates that encoding user preferences \textit{or} item attracting groups in the guidance signal is useful, rather than simply using node embeddings without including neighbor information.

\item Our complete model CG-KGR consistently performs the best, as it \textbf{jointly} encodes the information of user preferences and item attractions in the collaborative guidance signal to maximize the model performance. 
This shows the superior effectiveness and necessity of our proposed collaborative guidance mechanism, as it fully exploits the interactive semantics as information guidance in the listwise ranking tasks, i.e., Top-K recommendation.
\end{enumerate}

\subsubsection{\textbf{Case Study of Collaborative Guidance Mechanism.}}

\begin{figure}[tp]
\centering
\includegraphics[width=3.5in]{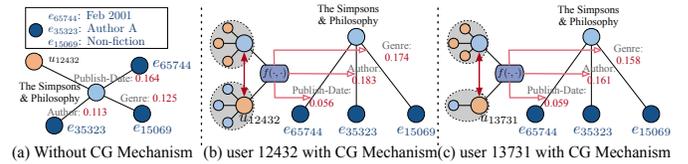}
\caption{A real example from Book dataset.}
\label{fig:case_study}
\end{figure}

To visualize the effect of the Collaborative Guidance Mechanism, we show a real case from Book dataset in Figure~\ref{fig:case_study}. 
As we can observe in Figure~\ref{fig:case_study}(a), without using Collaborative Guidance Mechanism, KG entities make similar contributions to the knowledge extraction with the weights 0.164, 0.125, and 0.113, respectively.
However, by using collaborative guidance based on the target pair $u_{12432}$ and book: \textit{The} \textit{Simpsons} \textit{\&} \textit{Philosophy}, our model can well customize the knowledge extraction:
CG-KGR highlights the knowledge triplets associated with entities $e_{35323}$ and $e_{15069}$ by calculating the weights as 0.183 and 0.174;
it pays less attention to the entity $e_{65744}$ with a lower weight 0.056.
This shows that, endorsed by our proposed Collaborative Guidance Mechanism, CG-KGR can well distinguish the informative knowledge (i.e., associated with relations \textit{Author} and \textit{Genre}) out of less informative one (i.e., associated with relation \textit{Publish\_Date}).
Futhermore, comparing Figures~\ref{fig:case_study}(b) with (c), we can find that users $u_{13731}$ and $u_{12432}$ show different collaborative influences on the customized knowledge extraction, as they have different focuses of item information.
Consequently, this is beneficial to personalized recommendation. 

\subsubsection{\textbf{Model Performance on Corrupted Book Dataset.}} 

\begin{figure}[h]
\centering
\includegraphics[width=3.5in]{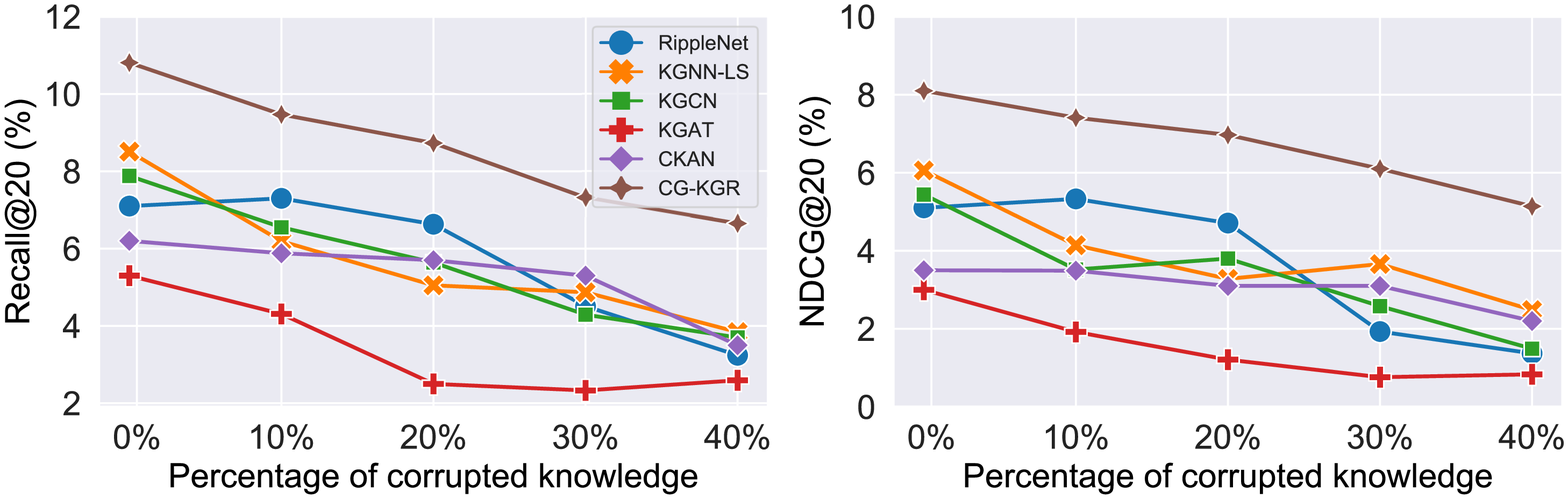}
\caption{Model performance on corrupted Book dataset.}
\label{fig:corrupt}
\end{figure}

We also conduct an interesting experiment on how Collaborative Guidance Mechanism defends the error/noise in KGs.
Specifically, we randomly generate corrupted knowledge in Book dataset and then replace it in the original KG: for example, we can replace a correct relation by a wrong one in the knowledge triplet.
The ratio of corrupted knowledge is ranging from 0-40\%.
We evaluate the performance of all KG-aware RS models on Top-K recommendation with corrupted information. 
As shown in Figure~\ref{fig:corrupt}, our proposed CG-KGR model can better defend the corrupted knowledge, with a Recall@20 decay from 10.81\% to about 6.65\%. 
By comparison, other models meet larger performance declines from about 8\% to 3\%. The reason of such phenomena is because they all lack a mechanism to sufficiently fuse internal interactive knowledge to guide the external information propagation. 
Meanwhile, our proposed Collaborative Guidance Mechanism lowers the negative influence of corrupt knowledge on the whole model learning and thus leads to better personalized recommendation.

\subsection{\textbf{Ablation Study of CG-KGR Model (RQ4)}}
To provide the intuition behind the performance improvement of CG-KGR, we conduct a comprehensive ablation study to evaluate the necessity of each model component.

\begin{table}[ht]
\centering
\caption{Ablation study on Top-20 recommendation (\%). }
\label{tab:ab_component}
\setlength{\tabcolsep}{1.4mm}{
\begin{tabular}{c| c | c | c | c |c}
\toprule
 Dataset      &   w/o EI &  w/o IL &  w/o ATT  & w/o CG & Best \\
\midrule
\midrule
        {\scriptsize MS-R@20}   & {\scriptsize 14.47}{\,{\tiny(-31.29\%)}}      
                                & {\scriptsize 20.76}{\,{\tiny(-1.42\%)}}      
                                & {\scriptsize 15.35}{\,{\tiny(-27.11\%)}}         
                                & {\scriptsize 18.53}{\,{\tiny(-12.01\%)}}  
                                &{\textbf{21.06}} \\
        {\scriptsize MS-N@20}   & {\scriptsize \ \,9.48}{\,{\tiny(-22.93\%)}}    
                                & {\scriptsize 12.13}{\,{\tiny(-1.38\%)}}      
                                & {\scriptsize 10.93}{\,{\tiny(-11.14\%)}}    
                                & {\scriptsize 11.10}{\ \,{\tiny(-9.76\%)}} 
                                &{\textbf{12.30}} \\   
 \midrule 
        {\scriptsize BK-R@20}   & {\scriptsize \ \,8.75}{\,{\tiny(-19.06\%)}}    
                                & {\scriptsize 10.32}{\,{\tiny(-4.53\%)}}   
                                & {\scriptsize \ \,9.78}{\,{\tiny(-9.53\%)}}
                                & {\scriptsize 10.44}{\ \,{\tiny(-3.42\%)}}  
                                &{\textbf{10.81}} \\
        {\scriptsize BK-N@20}   & {\scriptsize \ \,7.27}{\,{\tiny(-10.25\%)}}   
                                & {\scriptsize \ \,7.91}{\,{\tiny(-2.35\%)}} 
                                & {\scriptsize \ \,7.62}{\,{\tiny(-5.93\%)}}  
                                & {\scriptsize \ \,7.94}{\ \,{\tiny(-1.98\%)}} 
                                &{\ \,\textbf{8.10}} \\   
 \midrule
        {\scriptsize MV-R@20}   & {\scriptsize 19.86}{\,{\tiny(-20.40\%)}}       
                                & {\scriptsize 23.09}{\,{\tiny(-7.45\%)}}      
                                & {\scriptsize 24.01}{\,{\tiny(-3.77\%)}}     
                                & {\scriptsize 23.52}{\ \,{\tiny(-5.73\%)}}    
                                &{\textbf{24.95}} \\
        {\scriptsize MV-N@20}   & {\scriptsize 14.32}{\,{\tiny(-26.49\%)}}       
                                & {\scriptsize 18.51}{\,{\tiny(-4.98\%)}}       
                                & {\scriptsize 19.18}{\,{\tiny(-1.54\%)}}   
                                & {\scriptsize 18.39}{\ \,{\tiny(-5.60\%)}}   
                                &{\textbf{19.48}} \\   
 \midrule
        {\scriptsize RT-R@20}   & {\scriptsize 17.92}{\,{\tiny(-29.45\%)}}     
                                & {\scriptsize 23.74}{\,{\tiny(-6.54\%)}}    
                                & {\scriptsize 22.73}{\,{\tiny(-10.51\%)}}   
                                & {\scriptsize 21.19}{\,{\tiny(-16.57\%)}}     
                                &{\textbf{25.40}} \\
        {\scriptsize RT-N@20}   & {\scriptsize \ \,8.17}{\,{\tiny(-34.90\%)}}   
                                & {\scriptsize 11.72}{\,{\tiny(-6.61\%)}}     
                                & {\scriptsize 11.23}{\,{\tiny(-10.52\%)}}    
                                & {\scriptsize 10.85}{\,{\tiny(-13.55\%)}} 
                                &{\textbf{12.55}} \\   
\bottomrule
\end{tabular}}
\end{table}

\subsubsection{\textbf{Effect of Explicit Learning on Interactive Information.}}
To verify the effectiveness of explicit information learning for user-item interactions, we consider one variant of CG-KGR model by removing the interactive information propagation, which is denoted as CG-KGR$_{\rm w/o \, EI}$. 
As shown in Table~\ref{tab:ab_component}, variant CG-KGR$_{\rm w/o \, EI}$ remarkably underperforms CG-KGR.
This demonstrates that explicitly propagating interaction information is very important to boost CG-KGR performance.

\subsubsection{\textbf{Effect\,of\,Independently\,Learning\,on\,Two\,Data\,Sources.}}
We study the effect of independently learning user-item interaction and external knowledge by mixing the learning of these two parts together.
We denote this variant as CG-KGR$_{\rm w/o \, IL}$.
As shown in Table~\ref{tab:ab_component}, the variant CG-KGR$_{\rm w/o \, IL}$ confronts a conspicuous performance decay in recommendation, 
which justifies the effectiveness of distinguishing the learning processes of these two data sources in improving CG-KGR's model performance.

\subsubsection{\textbf{Effect of Knowledge-aware Attention Mechanism.}}
To substantiate the impact of our knowledge-aware attention mechanism, we use a variant, i.e., CG-KGR$_{\rm w/o \, ATT}$, by enabling the neighbors to equally contribute to the knowledge extraction. From Table~\ref{tab:ab_component}, we find that the results of CG-KGR$_{\rm w/o \, ATT}$ are worse than those of CG-KGR across all datasets. This supports that our knowledge-aware attention mechanism is effective to determine the knowledge informativeness in knowledge extraction phases, which finally leads to a big boost in Top-K recommendation task.

\subsubsection{\textbf{Effect of Collaborative Guidance Mechanism.}}
We disable Collaborative Guidance Mechanism by replacing $f(\boldsymbol{v}_u, \boldsymbol{v}_i)$ to an all-one vector in the follow-up model learning. 
This actually degrades our proposed quintuplet-based learning paradigm for knowledge extraction to the triplet-based, i.e., from $($$<$$u$,$i$$>$,$h,r,t)$ to $(h,r,t)$.
We denote it as CG-KGR$_{\rm w/o \, CG}$. 
As we can observe that, with all other model components, enabling our proposed Collaborative Guidance Mechanism can further improve the performance for personalized recommendation, showing that customized knowledge extraction is efficacious in improving CG-KGR performance.


\subsection{\textbf{Hyper-parameter Analysis (RQ5)}}

\subsubsection{\textbf{Implementation of Guidance Signal Encoder $f$.}}
We conduct experiments on different selections of the encoder $f$ and report the results in Table~\ref{tab:f}. From the results, while encoder $f_{pmax}$ works well on Music dataset, $f_{combine}$ shows consistent superiority over other selections on the other three datasets.  
$f_{combine}$ directly condenses information of user preferences and item attracting groups via the pairwise linear combining of embeddings, which is simple but effective in practice especially for these medium and large datasets. 

\begin{table}[h]
\centering
\caption{Top-20 recommendation of different $f$ (\%).}
\label{tab:f}
\setlength{\tabcolsep}{6.3mm}{
\begin{tabular}{c|c  c  c}
\toprule
        Dataset  & $f_{sum}$      & $f_{pmax}$  & $f_{comb}$\\
 \midrule
\midrule
        MS-R@20  &{18.07}       &{\textbf{21.06}}  &{20.61} \\
        MS-N@20  &{10.24}       &{\textbf{12.30}}  &{11.82}    \\
\midrule        
        BK-R@20  &{10.35}     &{\ \,9.28}&{\textbf{10.81}}        \\
        BK-N@20  &{\ \,8.07}  &{\ \,7.85}&{\ \,\textbf{8.10}}  \\
\midrule        
        MV-R@20  &{24.33}  &{23.49} &{\textbf{24.95}}       \\
        MV-N@20  &{18.50}  &{18.37} &{\textbf{19.48}}      \\
\midrule
        RT-R@20  &{21.84}  &{21.10} &{\textbf{25.40}}       \\
        RT-N@20  &{11.14}  &{11.33} &{\textbf{12.55}}       \\
\bottomrule

\end{tabular}}
\end{table}

\subsubsection{\textbf{Implementation of Information Aggregator $g$.}}

Most related recommendation work~\cite{KGCN,KGAT,RippleNet,CKAN} usually try  these aggregators and pick out the one with the best performance.
Hence, in this paper, we empirically report all performance likewise to explore the influence of aggregating neighbor information. 
As shown in Table~\ref{tab:agg}, under the scenario of Top-20 recommendation, $g_{concat}$ performs the best in general.
While for the Movie dataset, $g_{neighbor}$ surpasses the other two aggregators. 
This may be because $g_{neighbor}$ makes full use of external information by observing the entire neighborhood, which enlarges the predictive ranking power of CG-KGR model on Movie dataset. 

\begin{table}[h]
\centering
\caption{Top-20 recommendation of different $g$ (\%).}
\label{tab:agg}
\setlength{\tabcolsep}{5.8mm}{
\begin{tabular}{c|c  c  c}
\toprule
        Dataset  & $g_{sum}$      & $g_{concat}$  & $g_{neighbor}$\\
 \midrule
\midrule
        MS-R@20 &{15.61}    &{\textbf{21.06}}   &{17.01}\\
        MS-N@20 &{\ \,6.93}    &{\textbf{12.30}}   &{\ \,8.52}\\
\midrule        
        BK-R@20 &{10.54} &{\textbf{10.81}}   &{10.76}\\
        BK-N@20 &{\ \,7.83} &{\ \,\textbf{8.10}}&{\ \,8.03}\\
\midrule       
        MV-R@20 &{22.09}    &{20.13}   &{\textbf{24.95}}\\
        MV-N@20 &{16.31}    &{15.82}   &{\textbf{19.48}}\\
\midrule
        RT-R@20 &{21.95}    &{\textbf{25.40}}   &{20.86}\\
        RT-N@20 &{11.29}    &{\textbf{12.55}}   &{10.41}\\
\bottomrule

\end{tabular}}
\end{table}

\subsubsection{\textbf{Depth of Knowledge Extraction Hops.}}
\label{sec:layer}
We verify how the hop depth affects the performance by varying $L$ from 0 to 3, which depth 0 means no information aggregation from the knowledge graph side. For Top-K recommendation, CG-KGR achieves the best performance when $L$ is 1, 1, 2, and 1 for all benchmarks, respectively.  
This is because for Movie dataset, a relatively deeper knowledge extraction introduces more long-distance knowledge, which enriches the latent representation of items. 
As for the other three datasets, local knowledge in KGs is more informative for the training of the proposed model. 
In conclusion, preserving an appropriate depth of extraction hops can not only avoid the over-smooth problem~\cite{li2019deepgcns,li2018deeper} (details are referred in the time complexity analysis), but also enable maximized performance over different recommendation datasets. 

\begin{table}[t]
\centering
\caption{Top-20 recommendation of different $L$ (\%).}
\label{tab:layer}
\setlength{\tabcolsep}{4mm}{
\begin{tabular}{c|c c c c }
\toprule
 & {$L = 0$} & {$L = 1$} & {$L = 2$}  & {$L = 3$}  \\

\midrule
\midrule
        MS-R@20 &{18.25}       &{\textbf{21.06}}   &{16.54}   &{16.85}   \\
        MS-N@20 &{10.37}       &{\textbf{12.30}}   &{\ \,8.99}   &{\ \,8.43}   \\
\midrule
        BK-R@20 &{\ \,9.68}       &{\textbf{10.81}}   &{10.58}      &{9.79}   \\
        BK-N@20 &{\ \,7.51}   &{\ \,\textbf{8.10}}&{\ \,7.89}   &{\ \,7.67}   \\
\midrule 
        MV-R@20 &{19.84}       &{20.04}     &{\textbf{24.95}} &{22.41}  \\
        MV-N@20 &{15.26}       &{15.83}     &{\textbf{19.48}} &{17.62}  \\
\midrule
        RT-R@20 &{20.39}       &{\textbf{25.40}}  &{23.99}         &{23.45} \\
        RT-N@20 &{\ \,9.79}       &{\textbf{12.55}} &{12.13}          &{{11.81}} \\
\bottomrule
\end{tabular}}
\end{table}

\section{Related Works}
\label{sec:label}

Studying ubiquitous graph data has aroused interests in various applications~\cite{fang2017effective,IDX,zhang2019,defferrard2016} and incorporating KGs in recommender systems receives much attention recently.
Existing KG-aware RS models can be generally categorized into three branches: (1) path-based methods~\cite{yu2013collaborative,yu2014personalized,zhao2017meta,hu2018leveraging,shi2018heterogeneous}, (2) regularization-based methods~\cite{CKE,KGAT}, and (3) propagation-based methods~\cite{RippleNet,KGNNLS,KGCN,CKAN}:

\begin{itemize}[leftmargin=*]
\item \textbf{Path-based methods} leverage the connectivity patterns among items in KGs, i.e., meta-paths or meta-graphs, to provide additional guidance in the predictive model. 
Such meta-paths are generated by: (1) either defining constraint sub-patterns to concatenate the prominent paths~\cite{hu2018leveraging,yu2013collaborative}, (2) or relying on manual selection and path generation algorithms to directly find the targets~\cite{yu2014personalized,zhao2017meta,shi2018heterogeneous}.  The main inadequacies of path-based methods primarily lie that: defining effective sub-paths requires intensive input of domain knowledge and labor resources, which could be extremely expensive when the KG's are large-scaled and complicated. Furthermore, it is difficult to optimize the path retrieval for the recommendation, while the selected paths do a great impact on the final performance.
Thus in this paper, we exclude path-based methods for model comparison.


\item \textbf{Regularization-based methods} usually devise additional loss terms to capture the KG structures and fuse these to regularize the model training~\cite{KGAT,CKE,KTUP}. 
Based on the shared item embeddings, these methods merge the two tasks of general recommendation and KG completion to jointly train the model.
One deficiency is that all these regularization-based methods adopt a fixed term to control the regularization effect; however, in the whole training process, the two different training phases may not always make constant contributions. This implies that they may need more advanced strategies to determine the evolving values of regularization terms. 
Moreover, most regularization-based methods rely on traditional knowledge graph embedding methods to separately complete the KG training, while high-order semantic information in KGs and user-item interactions are not explicitly propagated, which may result in suboptimal representation learning for users and items.

\item \textbf{Propagation-based methods}, aiming at refining the entity representations, usually perform iterative information propagation under the graph convolutional/neural network framework for recommendation~\cite{zhang2021group,zheng2020price,fan2019graph,wu2019session,jin2020multi,chang2020bundle}.
KG-based propagation methods mainly focus on exploring KGs for information enrichment~\cite{RippleNet,KGNNLS,KGCN,CKAN}. With the auxiliary information passed along $l$-hop links in the KG, the embedding representations of users and/or items can be refined. After this feature propagation process, the final representation of an item is a mixture of its initial representation and information from its multi-hop neighbors. Based on the enriched embeddings, the user's preference towards candidate items can be more accurately predicted. 
Although many effective models have been proposed, the primary problem is that methods such as KGCN~\cite{KGCN}, KGNNLS~\cite{KGNNLS}, and RippleNet~\cite{RippleNet} only focus on propagating knowledge in the KG, but do not fully exploit the user-item interactions. This may lead to insufficient profiling for both users and items and thus the recommendation capability of RS models may be constrained. 
In addition, all these methods ignore the fact that these KGs are imported from external sources and may contain irrelevant information. Via limited volume for information propagation, uncorrelated and uninformative information may exclude the positive one, which suppresses the model performance. 
To address these issues, our CG-KGR model is proposed.
\end{itemize}

%

\section{Conclusion and Future Work}
\label{sec:con}
CG-KGR explicitly propagates collaborative information in user-item interactions to profile their latent representations. 
Based on this latent summarization, CG-KGR then seamlessly fuses this collaborative encoding as guidance to customize the knowledge extraction from external KGs. 
The extensive experiments well demonstrate that CG-KGR significantly improves the recommendation performance over baselines on both tasks of Top-K recommendation and Click-Through rate prediction.

As for future work, we point out two possible directions. (1) Unlike uniform neighbor sampling in this paper, we may explore a non-uniform sampler to screen out representative neighbors with high importance.
This may further improve the efficiency and effectiveness of KG-based recommender systems, especially for large-scale datasets. 
 (2) After the data integration of KGs and user-item interactions, the global data distribution may change. How to utilize such data distribution for better information propagation and aggregation is an important topic to investigate.



\bibliographystyle{IEEEtran}
\balance
\bibliography{ref}


 \end{document}